\documentclass[reprint, 10pt, aps, prx, superscriptaddress, floatfix]{revtex4-2}
\usepackage{amsmath}
\usepackage{graphicx} 
\usepackage[dvipsnames]{xcolor}
\usepackage[colorlinks, allcolors=NavyBlue]{hyperref}

\usepackage{natbib}
\usepackage{url}
\usepackage{textcase}
\usepackage{bm}
\newcommand{\change}[1]{{\color{black}{#1}}}

\begin{document}
\title{Optimal control of bit erasure in stochastic random access memory}
\author{Songela W. Chen}
\affiliation{Department of Chemistry, University of California, Berkeley, California 94720, USA}
\affiliation{Chemical Sciences Division, Lawrence Berkeley National Laboratory, Berkeley, California 94720, USA}
\author{David T. Limmer}
\email{dlimmer@berkeley.edu}
\affiliation{Department of Chemistry, University of California, Berkeley, California 94720, USA}
\affiliation{Chemical Sciences Division, Lawrence Berkeley National Laboratory, Berkeley, California 94720, USA}
\affiliation{Materials Sciences Division, Lawrence Berkeley National Laboratory, Berkeley, California 94720, USA}
\affiliation{Kavli Energy NanoScience Institute, Berkeley, California 94720, USA}
\date{\today}

\begin{abstract}
Energy costs of information processing are growing exponentially. Bit erasure is a key problem in this energy-information nexus, and a number of seminal relationships have been deduced regarding the relationship between thermodynamic costs and memory storage. To continue making progress in the modern era, however, requires confronting thermodynamic costs in realistic physical systems which operate away from equilibrium. Here, we explore the thermodynamic costs of bit erasure in a complementary metal oxide semiconductor model of two types of random access memory. We find dynamic random access memory dissipates the least amount of energy when operated in the quasistatic limit, where errors are also minimized. By contrast, static random access memory is most efficiently operated in finite time due to the energy required to maintain the state of the bit. We demonstrate a numerically robust optimization scheme using mean field theory and automatic differentiation, finding optimal protocols compatible with electrical engineering insights. These results provide a framework for operating realistic circuits in thermodynamically advantageous ways.
\end{abstract}

\maketitle

\section{Introduction}
With the explosive growth of information technology, data center electricity consumption is projected to double in the next five years \cite{energy_ai_2025}. In the context of the global energy crisis, it is crucial to confront physical costs of computation. 
Bit erasure is a well-studied problem in equilibrium thermodynamics, dating back to Landauer's insight that at least $k_\mathrm{B}T\ln 2$ of energy is required to erase one bit of information \cite{landauer_irreversibility_1961}. As such, it offers a benchmark for understanding the energetic costs of computation. Landauer's formulation and most subsequent studies, however, ignore the practical requirements of finite time computation and the effect of the physical substrate implementing erasure that is often operated within a nonequilibrium steady state \cite{sagawa_minimal_2009,berut_experimental_2012,wimsatt_refining_2021,dago_information_2021}. 
\change{Although some studies have confronted finite time bit erasure or considered nonequilibrium initial states, the majority of these results have limited applicability as they employ abstract confining potentials to model circuitry} \cite{PhysRevE.102.032105,zulkowski_optimal_2014, giorgini2023thermodynamic, boyd2022shortcuts,ciampini_erasure_2025,dago_reliability_2024}. Here, we find optimal tradeoffs between dissipated heat and accuracy of bit erasure in a realistic model of a logical circuit, kept out-of-equilibrium by voltage differences across transistors. Our results leverage contemporary machine-learning-based numerical optimization techniques and formal insights from stochastic thermodynamics to expand our understanding of the interplay between thermodynamics and information processing. More generally, our approach illustrates the application of optimal control techniques to nonequilibrium steady states.

The vast majority of information processing devices today use complementary metal-oxide-semiconductor (CMOS) technology \cite{taur_ning}. Until now, energy efficiency has improved by physically scaling down transistors on chips \cite{koomey2010implications}. However, we are approaching the nanoscale regime where component sizes are comparable to the mean free path of electrons, and thermal noise effects must be considered. This caveat renders traditional macroscopic models used for circuit design inappropriate. Several CMOS-inspired models have been proposed in recent years to account for energetic costs and noise effects in a way that is thermodynamically consistent \cite{gao_principles_2021, freitas_stochastic_2021, gu_counting_2020}, more realistically than the Gaussian noise assumptions of standard SPICE simulations in electrical engineering \cite{SPICE}. 
\change{While some prior studies probed reliability of information storage in these models \cite{helms_stochastic_2025,freitas_reliability_2022,murphy_dissipation-reliability_2026}, little work had investigated active erasure until recently \cite{shimizu_thermodynamic_2025,basile_learning_2024}. One outstanding question is the discrepancy between the observation in statistical physics that dissipation is minimized in the quasistatic limit, versus the electrical engineering conventional knowledge of an unavoidable accumulating power dissipation \cite{taur_ning}. We aim to bridge this gap in an architecture-specific manner using a stochastic thermodynamic framework.}

\begin{figure*}
    \centering
    \includegraphics[width=17.7cm]{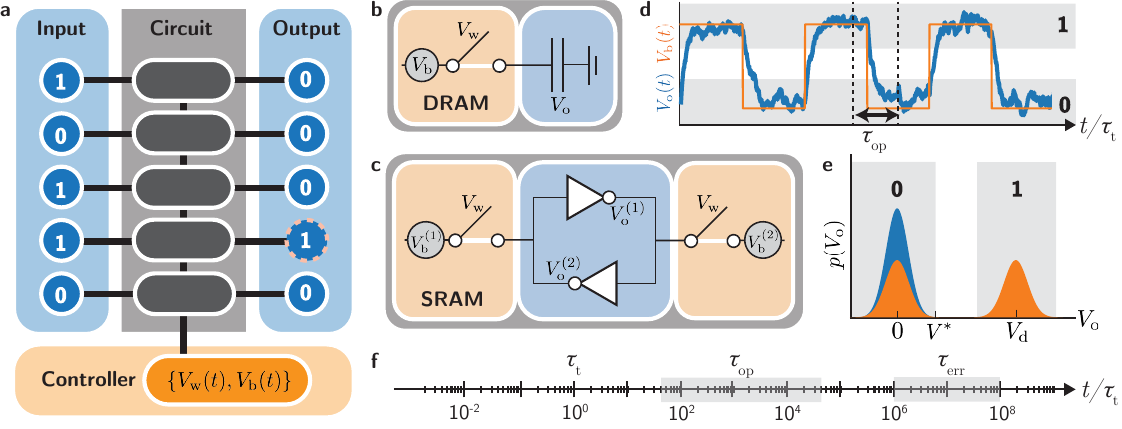}
    \caption{Schematic of bit erasure and related timescales. (a) Controlled bit erasure, driven by time-dependent $V_\mathrm{w}(t),V_\mathrm{b}(t)$ to set a 0 state. (b) Dynamic random access memory (DRAM) circuit. (c) Static random access memory (SRAM) circuit. (d) Idealized trajectory showing evolution of bit state $V_\mathrm{o}(t)$ (blue) in response to control $V_\mathrm{b}(t)$ (orange). (e) Probability distribution of bit state before (orange) and after (blue) erasure, with threshold voltage $V^*$ defining the 0 state. (f) Relevant timescales for electron residence time in a transistor $\tau_\mathrm{t}$, operation time $\tau_\mathrm{op}$, and stochastic error time $\tau_\mathrm{err}$. }
    \label{fig:general-characteristics}
\end{figure*}
In this study, we used one of these microscopic models to quantify the thermodynamic costs of bit erasure in a CMOS architecture of two different memory storage circuits, dynamic random access memory (DRAM) and static random access memory (SRAM). We use numerical optimization techniques to compare optimal erasure protocols, highlighting the contribution from the energy required to maintain the nonequilibrium steady states, as well as the reversible charging and discharging of the relevant capacitors. For CMOS circuits, bit erasure operates far from Landauer's bound, largely due to the sizable reversible heat from discharging the capacitors on which the bit is stored. Our results demonstrate that the tradeoffs between information processing speed and dissipation depend on circuit architecture. For SRAM, there is a thermodynamically optimal time for completing a bit erasure. By contrast, DRAM dissipates the least amount of energy and is most accurate when operated in the quasistatic limit. 

\section{Circuit model optimization}

We model CMOS implementations of memory storage devices using physical noise models consistent with circuits operated at low power \cite{gao_principles_2021, helms_stochastic_2025}. Here we study random-access memory (RAM), which is used to store working data rather than long-term data. 
In Fig.~\ref{fig:general-characteristics}(a), we show the general process.
The circuits that store bits of information are connected to so-called wordline $V_\mathrm{w}(t)$ and bitline $V_\mathrm{b}(t)$ voltages, which act jointly as a time-dependent controller that manipulates an output voltage $V_\mathrm{o}$ and sets it to 0 irrespective of the initial state. Individual bits are connected to these time-dependent controls through access transistors, which function as on-off switches. Macroscopically these are deterministic processes, but for systems operating close to thermal energy scales, fluctuations can result in errors in bit erasure and different amounts of heat dissipated in different realizations of erasure. We will consider averages over independent erasure attempts. 

In a typical CMOS circuit, like DRAM shown in Fig.~\ref{fig:general-characteristics}(b) or SRAM in Fig.~\ref{fig:general-characteristics}(c),  $V_\mathrm{w}$ controls the flow of current through the access transistor, and $V_\mathrm{b}$ controls the state being written. The basic unit of time is set by the residence time of an electron in a transistor, $\tau_\mathrm{t}$, while changes to $V_\mathrm{w}$ or $V_\mathrm{b}$ affecting bit erasure occur over times $\tau_\mathrm{op}\gg \tau_\mathrm{t}$, illustrated in Fig.~\ref{fig:general-characteristics}(d). The stochastic nature of the circuit is illustrated by the probability distribution before and after bit erasure in Fig.~\ref{fig:general-characteristics}(e), whereby errors can occur both spontaneously as well as the result of inaccurate control. The driving voltage $V_\mathrm{d}$ sets the scale of fluctuations and bias strength, and the threshold voltage $V^*$ defines the logical 0 state. The rate of spontaneous errors is denoted as $1/\tau_\mathrm{err}$, and as presented in Fig.~\ref{fig:general-characteristics}(f) we will consider operation timescales faster than it $\tau_\mathrm{op} \ll \tau_\mathrm{err}$ throughout.

\subsection{Stochastic circuit model} 
The components involved in either RAM device include electrodes, capacitors, and transistors. \change{Additional components such as inductors could also be accommodated for more complex devices.} We treat electrodes as ideal electron reservoirs at fixed potentials $V_j$. The capacitors are non-ideal electron reservoirs, with voltages across the capacitor $V_j$ that evolve dynamically. The charge on the capacitor is $q m_j = C V_j$, where $q$ is the fundamental unit of charge, $m_j=\{-\infty, \dots, 0, 1, \dots, \infty\}$ is the number of electrons on the capacitor, and $C$ is capacitance. In contrast to the electron reservoirs, we model transistors as a single Fermionic state, with occupancy $n_i = \{0, 1\}$ and onsite energy $E_i$ representing its band gap. The transistor energy is modulated electrostatically by the voltage sources to which it is connected. The state space of a given device is given by the joint occupancies of the capacitors and transistors, $\mathbf{x}=\{n_i\}^{N_\mathrm{t}} \otimes \{m_j\}^{N_\mathrm{c}}$ where $N_\mathrm{t}$ is the number of transistors and $N_\mathrm{c}$ the number of capacitors. Throughout, we use $i$ subscripts to reference transistor states and $j$ subscripts to reference electrode or capacitor states. We report voltages with reference to a thermal voltage $V_\mathrm{T}=26 ~\mathrm{meV}$ at room temperature and take capacitance $C=10 q/V_\mathrm{T}$.

The systems evolve stochastically according to the Markovian master equation
\begin{equation}
\partial_t \mathbf{p}(\mathbf{x}, t) = \mathbf{W p}(\mathbf{x}, t)
\end{equation}
where $\mathbf{p}(\mathbf{x})$ is the probability vector associated with the transistor and capacitor states and $\mathbf{W}$ is the stochastic generator, with matrix elements $W_{ij}$ specifying the rates of transition from states $j$ to $i$. 
The rates are defined with reference to the Fermi-Dirac distribution,
with the transition rate of an electron from an electrode $j$ to a transistor $i$ given by
\begin{equation}
    W_{ij}(V_j)=\Gamma \left(e^{\beta(E_i+q V_j)}+1\right)^{-1}   
\end{equation}
and the reverse given by
\begin{equation}
    W_{ji}(V_j)=\Gamma -\Gamma\left(e^{\beta(E_i+q V_j)}+1\right)^{-1}\, ,
\end{equation}
where $\beta=1/k_\mathrm{B}T$ is the inverse temperature defined with Boltzmann’s constant $k_\mathrm{B}$ and temperature $T$, 
$\Gamma$ specifies the timescale for transitions and is physically set by the resistance of the transistor-electrode interface, 
$E_i$ is a transistor energy, and 
$V_j$ is a fixed voltage measured on an electrode or the instantaneous voltage on a capacitor measured using a mid-point rule \cite{brillouin_can_1950}. 
We require $\Gamma \ll 1/\beta \hbar$ to ensure weak coupling such that electron hops can be treated as discrete  memoryless events, and simulation times are expressed in units of $\tau_\mathrm{t} = 1/\Gamma$ at room temperature.
 The  forward and reverse transition rates chosen in this way ensure local detailed balance and thus thermodynamic consistency \cite{limmer2024statistical}. This model is able to reproduce shot noise characteristics of CMOS devices \cite{sarpeshkar2002white}. We use kinetic Monte Carlo simulations to propagate the stochastic dynamics numerically \cite{gillespie_general_1976, gillespie_stochastic_2007}.

\subsection{Thermodynamic characteristics and optimization} 
To quantify thermodynamic costs of the bit erasure, we measure mean heat dissipation as the product of the electron current and its conjugate affinity from all channels in the circuit,
\begin{equation}
    \frac{d Q}{d t} 
    = \sum_{j ,i } J_{ji} \ln \frac{W_{ji}}{W_{ij}}
\end{equation}
where the sum is over the various pairs of electrodes and capacitors connected by individual transistors \cite{esposito_three_2010}.
For electrode or capacitor $j$ connected to transistor $i$, the current into electrode $j$ is given by
\begin{equation}
    J_{ji}( n_i) =  W_{ji} n_i  -W_{ij}(1 -  n_i ) \, ,
\end{equation}
and fluctuates due to the occupancy of the transistor and electron reservoir. 
To quantify the performance of the bit erasure, we define the error $\epsilon$ as the probability of being erroneously above a threshold voltage, $V^*$,
\begin{equation}
    \epsilon = \int_{V^*}^{\infty} dV_j \, p(V_j, \tau_\mathrm{op})
\end{equation}
where $p(V_j,\tau_\mathrm{op})$ denotes the probability density for $V_j$ of the measured output voltage at the final time $\tau_\mathrm{op}$. We will take the threshold to be $0.1V_\mathrm{d}$, \change{which is slightly looser than current practice at $V_\mathrm{d}=40 V_\mathrm{T} $\cite{gao_principles_2021}}.

To obtain optimal control protocols, we use automatic differentiation for nonequilibrium systems by Engel \textit{et al.} and the Adam optimization algorithm using the JAX library \cite{engel_optimal_2023, kingma_adam_2017, jax2018github, kidger2021on}. We define a loss function combining the averaged dissipated heat and error,
\begin{equation}
    \mathcal{L}=\langle Q \rangle+\lambda \epsilon
\end{equation}
where $\langle \dots \rangle$ denotes an average over independent realizations and $\lambda$ is a scaling factor for relative weights of the error and heat. As the error is bounded, we choose $\lambda$ large enough to ensure the minimal error at each control time. Details of the optimization may be found in Appendix \ref{opt-details}. Briefly, we use a nonlinear ansatz for the time dependences of both the wordline and bitline voltages and ensure that they are cyclical, $V_\mathrm{w}(0)=V_\mathrm{w}(\tau_\mathrm{op})$ and $V_\mathrm{b}(0)=V_\mathrm{b}(\tau_\mathrm{op})$  for operation time $\tau_\mathrm{op}$. Further, we restrict $V_\mathrm{w}=[-3V_\mathrm{d}, V_\mathrm{d}]$ and $V_\mathrm{b}=[0, V_\mathrm{d}]$, as values outside this range do not meaningfully change system dynamics and result in numerical instability. The coefficients in the control ansatz are initialized randomly and iterated stochastically in the direction of steepest descent. Rather than evaluating the loss and its gradient by propagating the master equation directly or for realizations of the stochastic dynamics, we use a surrogate model based on a mean field approximation to the stochastic dynamics described in Appendix \ref{mean-field-theory}.  \change{This surrogate model simplifies the gradient calculation greatly, making it comparable to previous continuous space algorithms that use genetic algorithms~\cite{whitelam2024nonequilibrium}.} As will be illustrated, the deterministic and fully differentiable mean-field dynamics are a very accurate representation of the typical trajectory of the system. Errors and mean dissipated heats are evaluated with the optimized protocols using kinetic Monte Carlo simulations. 

\section{Dynamic random access memory}

\begin{figure}
    \centering
    \includegraphics[width=8.5cm]{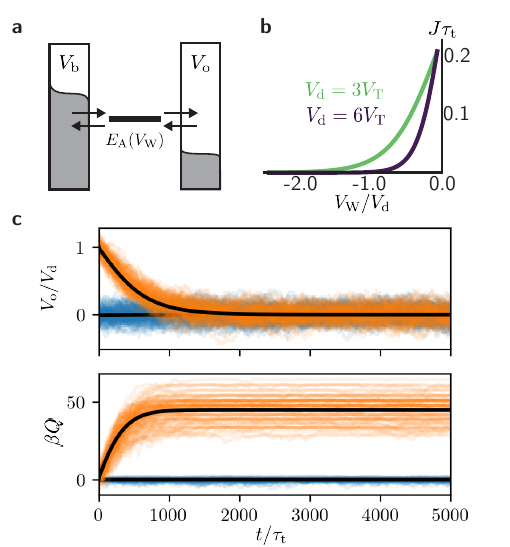}
    \caption{Dynamic random access memory (DRAM) phenomenology. (a) The DRAM circuit consisting of an ideal bitline electrode at voltage $V_\mathrm{b}$, an access transistor controlled by the wordline voltage $V_\mathrm{w}$, and a capacitor with output voltage $V_\mathrm{o}$ indicating the bit state. 
    (b) Average current across the transistor at constant values of $V_\mathrm{w}$ and different driving voltages $V_\mathrm{d}$. 
    (c) Example trajectories of bit erasure at $V_\mathrm{d}/V_\mathrm{T}=3$ starting in logical 1 (orange) or 0 (blue) compared to the estimate from mean field theory (black), for the output voltage (top) and heat (bottom).}
    \label{fig:dram-phenom}
\end{figure}

The two main types of RAM are dynamic (DRAM) and static (SRAM). 
 DRAM, shown in Fig.~\ref{fig:dram-phenom}(a), consists of an N-type access transistor connected to a bitline electrode at potential $V_\mathrm{b}$ and a storage capacitor with capacitance $C$. The transistor is also connected to a wordline electrode at potential $V_\mathrm{w}$, which controls the current through this transistor, functionally making it a switch. The onsite energy of the access transistor is given by $E_\mathrm{A}=q(V_\mathrm{b}-V_\mathrm{w})$. The voltage associated with charge on the capacitor serves as the output voltage, $V_\mathrm{o}$, which stores the state of the bit.
 The total energy of the system is a sum of the energy of the access transistor and the output capacitor
\begin{equation}
    E = n_\mathrm{A}q(V_\mathrm{b}-V_\mathrm{w}) + \frac{C V_\mathrm{o}^2}{2} \, 
\end{equation}
 where $n_\mathrm{A}$ is the occupancy of the access transistor and the output voltage $V_\mathrm{o} =  q m_\mathrm{o}/C$ is given by the occupancy of the capacitor. In between writing or erasing the bit, the bitline is held at an intermediate voltage, so that charge gradually leaks away from the capacitor over time. The leakage current depends on the energy of the transistor through both $V_\mathrm{w}$ and $V_\mathrm{b}$, as shown in  Fig.~\ref{fig:dram-phenom}(b). As a consequence of the leakage current, DRAM must be refreshed on timescales $\tau_\mathrm{err}$ longer than the bit erasure protocols, $\tau_\mathrm{op}$. \change{Given this separation of timescales, we do not consider the energetics of refreshing the memory in DRAM.}

Representative bit erasure trajectories are shown in Fig.~\ref{fig:dram-phenom}(c) starting from both 0 and 1 states. The two logical states are written into the capacitor through application of a bitline voltage $V_\mathrm{b}$ and setting the wordline voltage $V_\mathrm{w}=0$ to allow for current to flow. 
Writing the logical 1 state corresponds to setting $V_\mathrm{b}=V_\mathrm{d}>0$, while for the logical $0$ state, $V_\mathrm{b}=0$. 
Maintenance of the bit occurs by setting $V_\mathrm{b}=V_\mathrm{d}/2$ and $V_\mathrm{w}=-2V_\mathrm{d}$. As an example of a naive bit erasure protocol, analogous to that used in real CMOS device operation, at time $t=0$ the bitline voltage is set instantaneously to $V_\mathrm{b}=0$, and the wordline voltage is turned ``on" to $V_\mathrm{w}=0$ to allow current through the access transistor. The capacitor discharges with a characteristic RC time that is $\tau_\mathrm{RC} \sim  10^3 \tau_\mathrm{t}$, while dissipated heat accumulates in a manner that mirrors the voltage on the capacitor. Exact values of output voltages (top) and dissipated heat $Q$ (bottom) vary among stochastic trajectories, but their average is well described by a mean field approximation. We leverage the quantitative accuracy of this mean field theory to facilitate optimization. Starting in the 0 state, the average heat is 0, while starting in the 1 state, the average heat is given by the expected change in energy, $Q = -\Delta E \approx C V_\mathrm{d}^2/2$. Averaged over a uniform distribution of bits, the mean heat from this naive protocol would be $\langle Q\rangle  = -\Delta E \approx C V_\mathrm{d}^2/4$. 

\begin{figure}
    \centering
    \includegraphics[width=8.5cm]{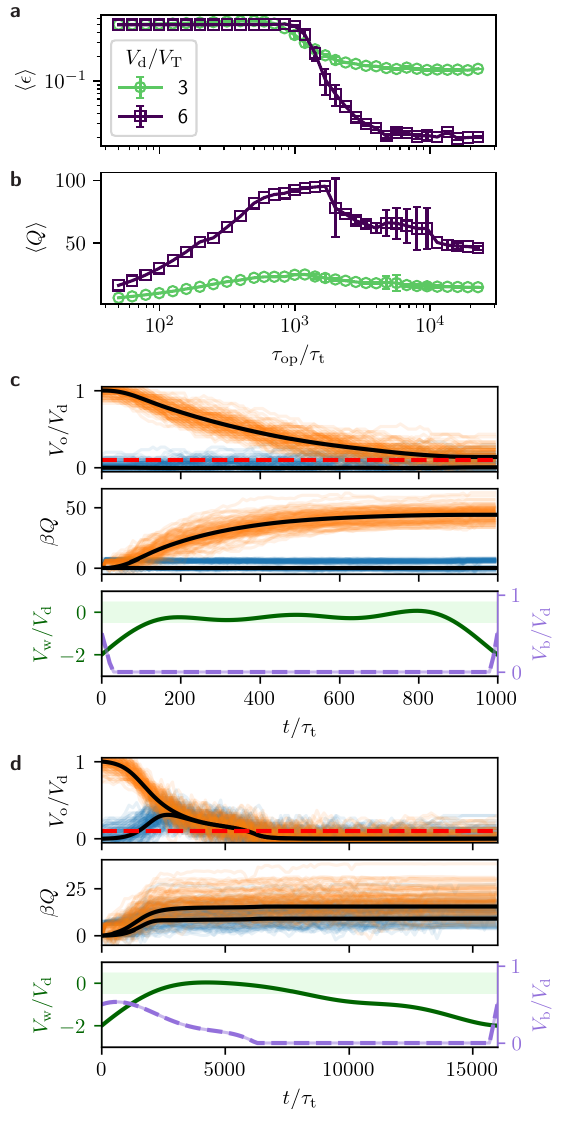}
    \caption{Heat-accuracy tradeoffs observed in DRAM, averaged over starting states 0 and 1. (a) Error and (b) dissipated heat for two driving voltages $V_\mathrm{d}/V_\mathrm{T}=3$ (green circles) and $V_\mathrm{d}/V_\mathrm{T}=6$ (purple squares) across operation times. Error bars indicate variation from degenerate protocols with optimization losses within 10\% of the minimum for that operation time $\tau_\mathrm{op}$.
    Representative trajectories and optimal protocols 
    \change{with $V_\mathrm{d}/V_\mathrm{T}=3$ for operation times (c) $\tau_\mathrm{op}/ \tau_\mathrm{t} =1000 $ and (d) $\tau_\mathrm{op}/\tau_\mathrm{t} =16000 $, respectively,} 
    where stochastic dynamics starting in logical 1 (orange) or 0 (blue) state are consistent with mean field dynamics (black). 
    \change{Red dashed line (top) indicates voltage threshold of $0.1 V_\mathrm{d}$. Green shaded region (bottom) indicates range of $V_\mathrm{w}$ where transistor is active.}}
    \label{fig:dram-tradeoffs}
\end{figure}

\begin{figure*}
    \centering
    \includegraphics[width=17.7cm]{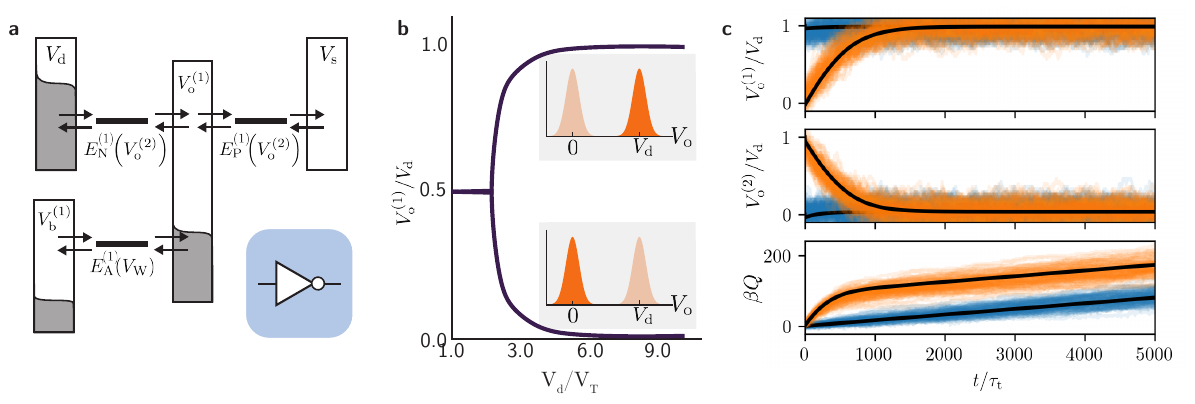}
    \caption{Static random access memory (SRAM) phenomenology. 
    (a) One of two coupled \textsc{not} gates in the SRAM circuit. It contains an ideal bitline electrode at $V_\mathrm{b}$, an access transistor controlled by the wordline  $V_\mathrm{w}$, and a capacitor with output voltage $V_\mathrm{o}$ indicating the bit state like in DRAM, but the capacitor is additionally connected to ideal source $V_\mathrm{s}$ and drain $V_\mathrm{d}$ electrodes with a nonequilibrium bias across them. 
    (b) Pitchfork bifurcation behavior as a function of driving voltage $V_\mathrm{d}$. Memory is stored when bistability emerges above a critical voltage $V_\mathrm{d}^* \approx 2.5$.
    (c) Example trajectories of bit erasure at $V_\mathrm{d}/V_\mathrm{T}=3$  starting in logical 1 (orange) or 0 (blue) compared to the estimate from mean field theory (black) for coupled output voltages (top, middle) and dissipated heat (bottom).
    }
    \label{fig:sram-phenom}
\end{figure*}
To understand the tradeoffs between dissipation and error, we have solved for optimal erasure protocols averaged over a uniform distribution of initial states of the bit 0 and 1. We consider the effects of both the operation time as well as the magnitude of the driving voltage, ensuring that the protocol is cyclic by enforcing $V_\mathrm{w}(0)=V_\mathrm{w}(\tau_\mathrm{op})$ and $V_\mathrm{b}(0)=V_\mathrm{b}(\tau_\mathrm{op})$. As a function of $\tau_\mathrm{op}$ we observe in Fig.~\ref{fig:dram-tradeoffs}(a) that at short operation times, $\tau_\mathrm{op} < \tau_\mathrm{RC}$, the error remains high, around 0.5. The source of the error is incoming bits that start in the 1 state and do not have time to relax fully to the logical 0 state. Beyond this time, the error rapidly decreases to a plateau value. The plateau error is inversely correlated with driving voltage $V_\mathrm{d}$, as larger $V_\mathrm{d}$ results in larger separation between the two equilibrium distributions for the two bit states $V_\mathrm{b}=V_\mathrm{d}$ and $V_\mathrm{b}=0$, so errors are determined by the persistence of thermal fluctuations. 
The average heat dissipation in Fig.~\ref{fig:dram-tradeoffs}(b) increases up to this minimum time required to erase the bit. Beyond this time, dissipation decreases, consistent with expectations from previous equilibrium models of finite time bit erasure. This is only possible by controlling both $V_\mathrm{w}(t)$ and $V_\mathrm{b}(t)$ with time. If we set $V_\mathrm{b}=0$ instantaneously as in standard CMOS operation, the minimum dissipation to achieve bit erasure is determined by the reversible charging energy of the capacitor $E=C V_\mathrm{d}^2/2$, and dissipation plateaus to 1/2 of this value in the long-time limit rather than to zero (see Appendix \ref{Vw-only}). Conversely, most prior studies on thermal noise in DRAM have manipulated the bitline voltage $V_\mathrm{b}$ while keeping the wordline voltage $V_\mathrm{w}$ constant \cite{van_brandt_non-landauer_2023, van_brandt_noisedissipation_2023, shimizu_thermodynamic_2025}, which is also slightly sub-optimal as will be shown. 

To understand the dependence of the heat and error on operation times, we consider the behavior of the circuit under limiting optimal protocols. In the short time regime, $\tau_\mathrm{op}<\tau_\mathrm{RC}$, optimal protocols resemble Fig.~\ref{fig:dram-tradeoffs}(c), where $V_\mathrm{b}$ and $V_\mathrm{w}$  are set to zero abruptly and return to their initial condition similarly sharply near $\tau_\mathrm{op}$. 
This protocol effectively turns on the access transistor by bringing the wordline voltage near 0 for the duration of the protocol, then turns it off at the end---just as typical devices do in practice. Physically, this corresponds to the energy of the access transistors being equal to the Fermi energy of their associated bitline electrodes. Optimal protocols have $V_\mathrm{w}$ slightly below 0, pulling the energy of the access transistor closer to that of the electron exiting the capacitor to facilitate more downhill transitions than having the access transistor resonant in energy with the bitline electrode only. Meanwhile, the bitline voltage $V_\mathrm{b}$ is set to the desired 0 state to maximize the thermodynamic force to discharge the logical 1 or preserve the logical 0. This is reflected in representative trajectories of $V_\mathrm{o}$ under this protocol, which either stay fluctuating near 0 from the start or immediately relax toward 0 once $V_\mathrm{w}$ reaches near 0. \change{Given the constraint of the short $\tau_\mathrm{op}$, the trajectories starting in 1 lack sufficient time to relax fully to 0, hence their end value is slightly above the threshold voltage $V^*=0.1V_\mathrm{d}$. However, we minimize the error by erasing the bit to the extent possible within this time and by ensuring incoming 0 states stay 0.} Mirroring the output voltage, the dissipated heat either fluctuates around 0 for the initial logical 0 state or grows as the capacitor discharges when starting in the logical 1. For simulations starting in a thermal distribution, the transistor is occupied roughly half the time, and the energy lost from this electron accounts for the small finite dissipation observed for some trajectories starting from the logical 0 state.

In the long-time limit, $\tau_\mathrm{op}>\tau_\mathrm{RC}$, optimal protocols resemble Fig.~\ref{fig:dram-tradeoffs}(d), where the access transistor is turned ``on" only for a brief amount of time while the bitline is adjusted quasistatically.
\change{From $t/\tau_\mathrm{t} \approx 1000$ to $t/\tau_\mathrm{t} \approx 6000$ in the example, $V_\mathrm{w}$ is near 0 so current can flow, and $V_\mathrm{b}$ decreases approximately linearly from $V_\mathrm{d}/2$ to 0.} 
During this active part of the protocol, the capacitor stays in equilibrium with the bitline voltage, minimizing the voltage difference across the transistor and therefore the dissipated heat. During the remainder of the protocol, the capacitor is in equilibrium with the bitline and so there are no additional channels for dissipation on average. Absent the initial logical 0 state, such a protocol would result in zero heat dissipated in the long time limit. However, we use a single protocol to erase from either 0 or 1 states, and a protocol that minimizes dissipation starting from the 1 state is necessarily non-optimal for the 0 state. The finite dissipation at long times reflect dissipation incurred from the 0 state moving away from 0 towards equilibrium with the bitline $V_\mathrm{b}$, and then returning back to 0 at the end of the protocol. Representative trajectories of $V_\mathrm{o}$ under this protocol starting in the logical 1 state exhibit a delay in their discharging before exhibiting an elongated relaxation to 0 due to the bitline being smoothly set to zero. In contrast, representative trajectories of $V_\mathrm{o}$ under this protocol starting in the logical 0 state exhibit a spike in values coincident with the turning on of the access transistor and bitline voltage at intermediate values, for example around $t/\tau_\mathrm{t}=2000$, before relaxing back to 0. The average heat dissipated for both initial bits mirror the output voltage, and for long operation times obtain a net contribution from the logical 0 state. An extremized version of this behavior is described in Appendix \ref{linear-response}. 

In both time limits, the basic phenomenology we find is independent of the $V_\mathrm{d}$ value we use to set the logical 1 state, though quantitatively the asymptotic error is strongly dependent on its value and the heat dissipated at finite time grows as $V_\mathrm{d}^2$. In the long time limit, the heat dissipated will approach $CV_\mathrm{d}^2/4$ or half the reversible charging energy of the capacitor, far greater than Landauer's bound. In principle there are contributions to the dissipated heat from the manipulation of the transistor energy, upon which work is being done. However, the characteristic relaxation time of the transistor, $\tau_\mathrm{t} \ll \tau_\mathrm{op}$, is much less than the protocol times we consider here. Thus, manipulation of the transistor maintains adiabaticity with respect to the protocol.

\section{Static random access memory}

In contrast to the 1-transistor, 1-electrode, 1-capacitor design of the DRAM, the 6-transistor, 6-electrode, 2-capacitor SRAM we consider next is more complex. This device consists of two coupled \textsc{not} gates, the kinetic network for which is shown in Fig.~\ref{fig:sram-phenom}(a). Each \textsc{not} gate has a source voltage, $V_\mathrm{s}=0$, a drain voltage, $V_\mathrm{d}>0$,  a control voltage electrode held at $V_\mathrm{b}^{(k)}$, and a capacitor whose voltage serves as the readout of the bit, $V_\mathrm{o}^{(k)}$, where $k$ labels the two \textsc{not} gates, $k=1,2$. Each \textsc{not} gate consists of an N-type and P-type transistor, which we model as a single state. The onsite energy of each transistor is determined by the output voltage of the other \textsc{not} gate, for example
\begin{align}
    E_\mathrm{P}^{(1)} =  qV_\mathrm{o}^{(2)}\, , \quad E_\mathrm{N}^{(1)} = \frac{3}{2} q V_\mathrm{d} - qV_\mathrm{o}^{(2)} \, ,
\end{align}
with occupation variables $n_\mathrm{P,1}$ and $n_\mathrm{N,1}$, and equivalent energies for the transistors on the second \textsc{not} gate, dependent on the output voltage of the first. This model of a \textsc{not} gate is able to reproduce expected voltage transfer curves in typical CMOS circuits \cite{gao_principles_2021}. The interdependency of the two \textsc{not} gates forms a positive feedback loop that maintains the state of the bit through a pitchfork bifurcation, exhibited in Fig.~\ref{fig:sram-phenom}(b). The bistability depends on the size of the driving voltage, $V_\mathrm{d}$, with an onset of bistability at $V_\mathrm{d}/V_\mathrm{T}\approx 2.5$, and a time to spontaneously flip a bit, $\tau_\mathrm{err}$, that grows exponentially with $V_\mathrm{d}$ \cite{helms_stochastic_2025}. \change{The combined output of the two capacitors determines the state of the bit, and we use the convention that $(V_\mathrm{o}^{(1)} > 0.9 V_\mathrm{d}) \land (V_\mathrm{o}^{(2)}< 0.1V_\mathrm{d})$ defines the 0 state.}

To control the state of the bit, each \textsc{not}-gate capacitor is  connected to an access transistor whose energies depend on a joint wordline voltage $V^{(1)}_\mathrm{w}=V^{(2)}_\mathrm{w}=V_\mathrm{w}$ and complementary bitline voltages $V_\mathrm{b}^{(1)}=V_\mathrm{b}$ and $V_\mathrm{b}^{(2)}=V_\mathrm{d}-V_\mathrm{b}$, with onsite energies $E_\mathrm{A}^{(k)}=q(V_\mathrm{b}^{(k)} - V_\mathrm{w})$. The current behavior shown in Fig.~\ref{fig:dram-phenom}(b) for DRAM also holds for the access transistors in SRAM. The resulting energy for the SRAM device is
\begin{equation}
E = \sum_{k=1,2} n_{\mathrm{P},k} E_\mathrm{P}^{(k)} +n_{\mathrm{N},k} E_\mathrm{N}^{(k)}  + n_{\mathrm{A},k} E_\mathrm{A}^{(k)}+  \frac{C}{2} ( V^{(k)}_\mathrm{o})^2
\end{equation}
where the output voltages are dynamic variables, determined by occupation variables $ m^{(k)}_\mathrm{o}$, as  $V^{(k)}_\mathrm{o} =  q m^{(k)}_\mathrm{o}/C$.

\begin{figure}
    \centering
    \includegraphics[width=8.3cm]{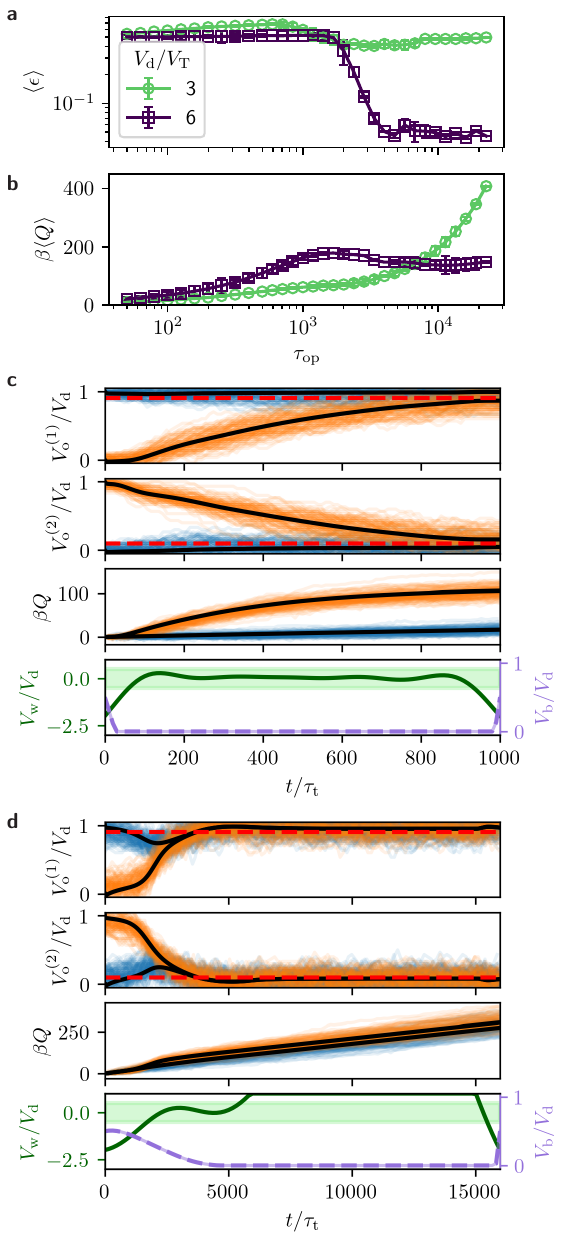}
    \caption{Heat-accuracy tradeoffs observed in SRAM, averaged over starting states 0 and 1. (a) Error and (b) dissipated heat for two driving voltages $V_\mathrm{d}/V_\mathrm{T}=3$ (green circles) and $V_\mathrm{d}/V_\mathrm{T}=6$ (purple squares) across operation times. Error bars indicate variation from degenerate protocols with optimization losses within 10\% of the minimum for that operation time $\tau_\mathrm{op}$.
    Representative trajectories and optimal protocols
     \change{with $V_\mathrm{d}/V_\mathrm{T}=3$ for operation times (c) $\tau_\mathrm{op}/ \tau_\mathrm{t} =1000 $ and (d) $\tau_\mathrm{op}/\tau_\mathrm{t} =16000 $, respectively,} 
    where stochastic dynamics starting in logical 1 (orange) or 0 (blue) state are consistent with mean field dynamics (black). 
    \change{Red dashed line (top) indicates voltage threshold of $0.1 V_\mathrm{d}$. Green shaded region (bottom) indicates range of $V_\mathrm{w}$ where transistor is active.}}

    \label{fig:sram-tradeoffs}
\end{figure}

SRAM is an intrinsically nonequilibrium system because the source and drain electrodes for each \textsc{not} gate are held at different voltages. Representative erasure trajectories are shown in Fig.~\ref{fig:sram-phenom}(c) starting from both 0 and 1 states. As with DRAM, two logical states are written into the capacitor by applying a bitline voltage and turning on the wordline voltage $V_\mathrm{w} = 0$ to allow current to flow across the access transistor. Writing the logical 1 state corresponds to setting $V_\mathrm{b} = V_\mathrm{d} > 0$, while for the logical 0, $V_\mathrm{b} = 0$. Maintenance of the bit occurs by setting $V_\mathrm{b} = V_\mathrm{d}/2$ and $V_\mathrm{w} = -2V_\mathrm{d}$. Like DRAM, bit erasure for SRAM occurs with a naive protocol and with an analogous characteristic time $\tau_\mathrm{RC}$. However, the total dissipation in SRAM continues to increase after bit erasure has occurred. This so-called housekeeping heat \cite{seifert_stochastic_2012} contribution derives from the constant voltage drop across each \textsc{not} gate and changes the heat-accuracy tradeoffs for SRAM as compared to DRAM. The mean field theory captures this housekeeping heat contribution alongside the overall average behavior with quantitative accuracy, justifying its use in optimization.

To understand the dependence of the heat and error on operation time, we again consider the behavior of the circuit under limiting optimal protocols.
In Fig.~\ref{fig:sram-tradeoffs}(a), we show error $\epsilon$ as a function of operation time $\tau_\mathrm{op}$ at two driving voltages $V_\mathrm{d}/V_\mathrm{T}=3$ and 6. 
For operation times shorter than a threshold time, the error remains high, while at long times the error drops to a plateau value. Both the minimum time to erase and the plateau error rate depend on the driving voltage, with smaller $V_\mathrm{d}$ requiring shorter times but yielding larger errors asymptotically. The origin of the error in the long time limit is analogous to DRAM where smaller driving voltages produce larger overlap between the stable bit states, allowing thermal fluctuations to weakly degrade memory. The $V_\mathrm{d}$ dependence to the threshold time to reach this plateau error is a manifestation of the increase in stability of the two logical states inherited from the bifurcation in SRAM. Larger $V_\mathrm{d}$ results in deeper metastability of both the logical 0 and logical 1, and thus a longer time to transition between them \cite{helms_stochastic_2025}. 

In Fig.~\ref{fig:sram-tradeoffs}(b), we show the operation time dependence of the dissipated heat $Q$. Below the threshold time required to erase the bit, the average heat dissipated grows weakly because little activity occurs in the circuit. Dissipation peaks near the threshold time, then slowly decreases as operation time increases like in DRAM. However, for SRAM dissipation increases in the long-time limit due to the housekeeping heat, which results from stochastic transitions generating a continuous electron current from source to drain in each \textsc{not} gate. The housekeeping heat is more significant for lower driving voltages because these stochastic transitions are more likely across smaller potential differences, generating larger leakage currents. Thus, the SRAM heat-accuracy tradeoff becomes more dramatic for low driving voltage $V_\mathrm{d}/V_\mathrm{T}=3$ in the long-time limit, as housekeeping heat is a bigger contribution. As a consequence, optimal driving voltages depend on the desired operation time.

Representative optimal protocols are shown for two different $\tau_\mathrm{op}$ in Fig.~\ref{fig:sram-tradeoffs}(c-d). For short times in Fig.~\ref{fig:sram-tradeoffs}(c), current is maximized through the access transistors by setting the wordline voltage $V_\mathrm{w}$ near 0, strongly driving the system toward the zero state induced by setting the bitline voltage to $V_\mathrm{b}=0$. Error and dissipation are therefore minimized within the time constraint, the latter because energy is directed efficiently toward bit erasure rather than through other parts of the circuit. By contrast, a long-time protocol in Fig.~\ref{fig:sram-tradeoffs}(d) maintains the access transistors in ``off" states for most of the protocol. During the ``active" portion of the protocol where $V_\mathrm{w}\approx 0$, denoted by the shaded regions, $V_\mathrm{b}$ follows the energy of the discharging capacitor within the circuit. In this case, it is advantageous to minimize activity in the circuit during the ``off" phase to minimize dissipation, while still ensuring that the bit is erased during the subsequent ``on" phase. We found that the duration of the ``on" part of the protocol could vary somewhat without changing the results dramatically, and also that the exact protocol during the ``on" part could vary as long as $V_\mathrm{w}$ was near 0. Accordingly, the behavior of $V_\mathrm{b}$ while $V_\mathrm{w}$ is away from zero does not affect the state of the bit. Various optimal protocols are nearly degenerate, determined by judicious switching between the ``off" and ``on" phases of the control protocol.

In both limits of $\tau_\mathrm{op}$, the optimal protocol resembles that of DRAM because the same control parameters $V_\mathrm{w}(t)$ and $V_\mathrm{b}(t)$ are tuned on timescales much slower than the characteristic relaxation time of the transistor, $\tau_\mathrm{op} \gg \tau_\mathrm{t}$. Work done to manipulate the energy of the occupied transistor is negligible as compared to the reversible charging energy of the capacitor and especially to the housekeeping heat. Unlike DRAM, however, the dissipation is not minimized in the long time limit and grows unbounded due to the housekeeping heat. This time-extensive contribution was also observed in a similar study using an alternative thermodynamically consistent model of SRAM \cite{basile_learning_2024}. The optimal time to perform erasure depends on $V_\mathrm{d}$ and is coincident with the onset of the plateau in the error. For large $V_\mathrm{d}$ there is a range of optimal operation times $\tau_\mathrm{op} > \tau_\mathrm{RC}$, as the timescale to plateau the error and that associated with the growth of housekeeping heat are distinct, yielding an optimal dissipated heat on average of $C V_\mathrm{d}^2$. For small $V_\mathrm{d}$ there is less of a separation of timescales between the plateau of the error and the growth of housekeeping heat, so there is a narrow range of operation times able to reach this limiting heat. The basic phenomenology is reproduced with simple linear protocols motivated from the study of DRAM, illustrated in Appendix \ref{linear-response}.

\section{Discussion}
We have shown that physical details are important when accounting for thermodynamic costs of bit erasure. For DRAM \change{operated on timescales shorter than the memory refresh time}, we find that a quasistatic protocol is optimal, in line with prior studies that assume equilibrium thermodynamics \cite{sagawa_minimal_2009, zulkowski_optimal_2014, proesmans_finite-time_2020}. For SRAM, we find an optimal tradeoff between error and dissipated heat at intermediate operation times due to the accumulation of housekeeping heat. 
\change{This contribution from housekeeping heat has been recently pointed out in Ref.~\cite{basile_learning_2024}, but it was neglected in prior studies that work with simple potentials, even though background power consumption is well studied in electrical engineering \cite{taur_ning}. As a consequence, we have laid a foundation for significant further study on the thermodynamics of modern computation.}

By understanding the behavior of our time-dependent control protocols, we obtain a roadmap for future design principles in nanoscale devices using similar architectures.
\change{The approach we have developed is distinct from previous calculations that optimize erasure using the master equation directly~\cite{basile_learning_2024}, which while providing easy access to gradients is limited in system size. By using the kinetic Monte Carlo method in conjunction with gradients approximated from mean field theory, we have a methodology that can scale to large systems but is still optimizable.} Given this numerically robust and procedurally flexible nonequilibrium optimization scheme, we hope to extend this framework to more complex circuits with larger-scale implications. \change{Future work will help to elucidate other energetically costly operations used in contemporary CMOS computers, and hopefully help to attenuate the increasing energy demands of information processing technology. }

\section*{Data Availability}
The source code for the calculations done and all data presented in this work 
are openly available \cite{code}.

\section*{Acknowledgments}
We thank Gavin Crooks, Sam Oaks-Leaf, Kyle Ray, and Michael DeWeese for helpful discussions.
This work was supported by NSF Grant CHE-1954580 and by the U.S. Department of Energy, Office of Science, Office of Basic Energy Sciences, Materials Sciences and Engineering Division, under Contract No. DEAC02-05-CH11231 within the Fundamentals of Semiconductor Nanowire Program (KCPY23). This research used the Savio computational cluster resource provided by the Berkeley Research Computing program at the University of California, Berkeley (supported by the UC Berkeley Chancellor, Vice Chancellor for Research, and Chief Information Officer).

\appendix

\section{Mean field theory} \label{mean-field-theory}
To enable the use of automatic differentiation and to accelerate numerical optimization, we use a mean field theory to propagate the various components. We assume that transition rates depend on the mean value of each fluctuating quantity, 
such that we propagate coupled ordinary differential equations for the equations of motion of each component, 
\begin{gather}
    \frac{d \langle n_i \rangle}{d t} = \sum_j J_{ij} (\langle m_j\rangle, \langle n_i\rangle )\\
    \frac{d \langle m_j \rangle}{d t} = \sum_i J_{ji}(\langle m_j\rangle, \langle n_i\rangle)\,
\end{gather}
where the sum goes over the connected components.
Consequently, we obtain 2 coupled ordinary differential equations for DRAM, with access transistor occupancy $n_A$ and capacitor state $m_o$,
  and 8 coupled ODEs for SRAM, including N-type, P-type, and access transistor occupancies $n_i^{(k)}$, 
 and output capacitor state $m_o^{(k)}$ for each of $k=1,2$ \textsc{not} gates.

This mean field theory is accurate in the macroscopic limit, where individual electronic transitions become uncorrelated relative to the overall scale of the system. In practice we find that working capacitances of $C=10$ and driving voltages $V_\mathrm{d}/V_\mathrm{T}=3,6$ are sufficient to yield quantitatively accurate results for our system as shown in the main text. For this system, fluctuations are roughly symmetry about the mean and therefore the theory works well. If the distribution were more skewed or we were interested in information about the tails of the distribution, a more sophisticated theory capturing the full trajectory dynamics would be necessary.
Using this framework, we propagate mean-field dynamics within the optimization process for numerical facility, then confirm the results using kinetic Monte Carlo simulations with the optimized parameters.

\section{Optimization details} \label{opt-details}
\begin{figure}
    \centering
    \includegraphics{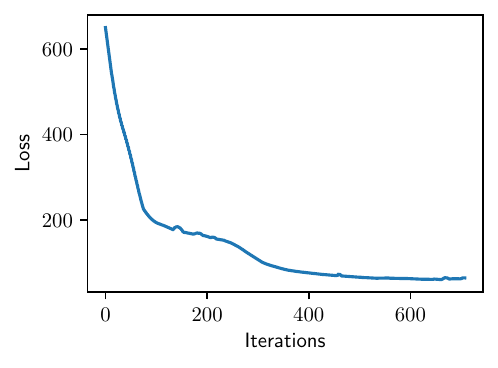}
    \caption{Loss function starting from random seed.}
    \label{fig:loss}
\end{figure}

In a high-dimensional and non-convex parameter space subject to nonequilibrium driving, finding an optimal solution is difficult. Engel et al. recently developed a method to find optimal protocols for nonequilibrium systems using automatic differentiation \cite{engel_optimal_2023}. We use this method with our fully differentiable mean field theory to propagate dynamics. For our bit erasure setup, we found that a learning rate of $10^{-3}$ achieved sufficient convergence within 1000 iterations of the Adam optimizer with otherwise standard parameters. We used $\lambda=10$ to scale the relative contributions in our loss function, chosen empirically to ensure complete bit erasure while still allowing a dynamic range of heat dissipation to influence the optimization. 
We used several different initializations to generate optimal protocols, then used the best results. First, for each length of protocol, we optimized 10 sets of randomly initialized seed coefficients and took the best of these 10 as the optimal protocol. An example of the convergence of the optimization loss is shown in Fig.~\ref{fig:loss}, and a successful erasure protocol was obtained. However, the roughness of the optimization landscape created inconsistencies in optimal protocols across parameter spaces.

Noticing two classes of solutions from the random seeds at short and long times, we used the shortest time optimal protocol to seed the next shortest time optimization iteratively, and the same from the longest time protocol. There was significant hysteresis in the optimal protocols obtained from these two sequential seeds.
Finally, we took the long-time-seeded protocols and refined the optimization for each time point by seeding from time points slightly shorter and longer than the target operation time. The refined protocols are used for the data in the main text.

\section{Restricted optimization} \label{Vw-only}

\begin{figure}
    \centering
    \includegraphics[width=8cm]{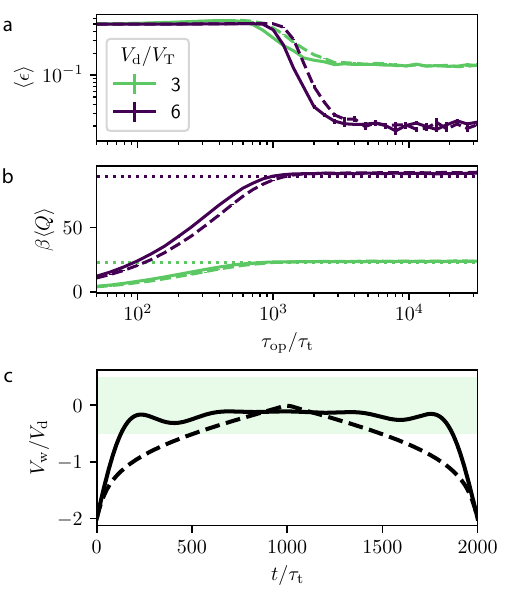}
    \caption{Heat-accuracy tradeoffs observed in DRAM with only wordline $V_\mathrm{w}$ control, averaged over starting states 0 and 1. (a) Error and (b) dissipated heat for two driving voltages $V_\mathrm{d}/V_\mathrm{T}=3$ (green) and $V_\mathrm{d}/V_\mathrm{T}=6$ (purple) across operation times. 
    (c) Representative optimal protocol, similar for all operation times.
    Solid lines denote data from automatic differentiation, and dashed lines are the same from the linear response theory. Dotted horizontal lines show $C V_d^2/4$, half the energy of a capacitor. }
    \label{fig:dram-tradeoffs-vw-only}
\end{figure}

In a standard CMOS device undergoing bit erasure, the bitline is precharged to $V_\mathrm{b}=0$ and the wordline $V_\mathrm{w}$ is turned on or off. We performed a single parameter optimization of the wordline only to probe the effects of modulating current through the access transistor. The overall takeaways hold from the 2-dimensional optimization: that there is a minimum time required for erasure, that error plateaus for times longer than the minimum, and that there is a tradeoff between error and dissipation. In Fig.~\ref{fig:dram-tradeoffs-vw-only}(a), we show for DRAM the same plateau of error that was observed in the optimization with both $V_\mathrm{b}$ and $V_\mathrm{w}$, and that error is inversely related to driving voltage. However, given a fixed bitline voltage $V_\mathrm{b}=0$, dissipation also plateaus to the reversible charging energy of the capacitor, $E=CV_\mathrm{d}^2/2$, rather than decreasing in the long-time limit in Fig.~\ref{fig:dram-tradeoffs-vw-only}(b).

Optimal protocols for this constrained case shown in Fig.~\ref{fig:dram-tradeoffs-vw-only}(c) are remarkably simple. There is no opportunity to adjust the bitline voltage adiabatically, so any electron exiting the system goes to a bath at potential $V_\mathrm{b}=0$. The only adjustable behavior is flux through the access transistor, and since the system is relaxing to an equilibrium state, we need only to turn on the access transistor to allow the relaxation process to occur. Hence, the protocol stays near $V_\mathrm{w}=0$ for most of the time, still respecting the boundary conditions where it is in the ``off" state at $V_\mathrm{w}(0)=V_\mathrm{w}(\tau_\mathrm{op})=-2V_\mathrm{d}$.

Tradeoffs and optimal protocols were also computed for SRAM in this restricted framework. At fixed bitline voltage $V_\mathrm{b}=0$, the minimum dissipation is the reversible charging energy of the two output capacitors, $E=CV_\mathrm{d}^2$. As a nonequilibrium system, SRAM incurs predominantly housekeeping heat at long times. Optimal protocols here are also controlling only flux through the access transistor. We cannot simply turn on the access transistor in SRAM as we did in DRAM because it is not simply an equilibrium relaxation process. However, the behavior is similar to that of the optimization with both $V_\mathrm{b}$ and $V_\mathrm{w}$ for SRAM, where short operation times minimize error by maximizing the discharging flux through the access transistor, and long operation times minimize dissipation by minimizing overall activity.

\section{Linear response} \label{linear-response}
Some insight into the optimal protocols can be gained using linear response theory \cite{sivak_thermodynamic_2012}, applied in the limit of long operation times where erasure is assumed and the dissipated heat minimized. Such a theory is applicable to equilibrium systems, so we apply it to DRAM in which maintenance of the bit is not considered, only the erasure, and leakage current is ignored. Under these assumptions for a protocol that controls both wordline and bitline voltages, $\mathbf{V}(t)=\{V_\mathrm{w}(t),V_\mathrm{b}(t)\}$, the dissipation is given by
\begin{equation}
\langle Q \rangle = \int_0^{\tau_\mathrm{op}} dt\, \frac{d\mathbf{V}}{dt} \cdot \boldsymbol{\zeta} \cdot  \frac{d\mathbf{V}}{dt}
\end{equation}
where $\boldsymbol{\zeta}$ is a friction matrix, given by a matrix of integrated time correlation functions. In the case of DRAM, the relevant matrix has entries that only depend on the autocorrelation function of the occupancy of the access transistor, as the driving force is solely a function of $n_A$. It follows from
\begin{equation}
   -\frac{\partial E}{\partial \mathbf{V}}=\{ n_A, -n_A \}
\end{equation}
that the friction matrix $\boldsymbol{\zeta}$ is 
\begin{equation}
  \boldsymbol{\zeta} = \begin{bmatrix} 1 & -1 \\ -1 & 1 \end{bmatrix} \zeta \, ,
\end{equation}
where
\begin{align}
    \zeta &= \beta \int_0^\infty dt \, \langle \delta n_A(0) \delta n_A(t) \rangle =\frac{\beta}{ 4\Gamma(1+\cosh(\beta q V_\mathrm{w}))} 
\end{align}
is a correlation function that can be evaluated exactly for DRAM by solving for the average occupancy in equilibrium since $ \langle \delta n_A^2\rangle=\langle n_A\rangle (1-\langle n_A \rangle )$.

The friction matrix is thus a function of $V_\mathrm{w}$ but independent of $V_\mathrm{b}$. As a consequence, optimal protocols for $V_\mathrm{b}$ are piecewise linear in time, while those for $V_\mathrm{w}$ satisfy an Euler-Lagrange equation of the form
\begin{equation}
\frac{d^2V_\mathrm{w}}{dt^2} = \frac{\beta q \sinh(\beta q V_\mathrm{w})}{2+2 \cosh(\beta q V_\mathrm{w})} \left( \frac{d V_\mathrm{w}}{dt}\right )^2
\end{equation}
with boundary conditions $V_\mathrm{w}(0)= -2V_\mathrm{d}$ and $V_\mathrm{w}(\tau_\mathrm{op}/2)= 0$, corresponding to the ``off" and ``on" states of the access transistor, respectively. As a controlled bit erasure requires the same start and end states, we symmetrize the optimal protocol to return to $V_\mathrm{w}(\tau_\mathrm{op})=-2V_\mathrm{d}$.

Considering first only controlling $V_\mathrm{w}$, although this analytical form cannot predict the optimal length of protocol $\tau_\mathrm{op}$, nor can it tell us how long we should keep the transistor ``on,"  it informs us about the shape of the optimal protocol, shown in the dashed line in Fig.~\ref{fig:dram-tradeoffs-vw-only}(c). We observe the same rapid increase in the control parameter as obtained by automatic differentiation, followed by a slower transition towards $V_\mathrm{w}=0$. The corresponding thermodynamic characteristics in dashed lines in Figs. \ref{fig:dram-tradeoffs-vw-only}(a) and (b) show that this protocol performs  comparably to the optimal protocol from automatic differentiation, only slightly worse quantitatively.

\begin{figure}
    \centering
    \includegraphics[width=6cm]{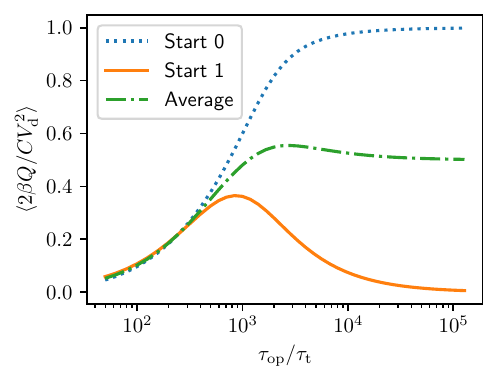}
    \caption{Dissipated heat for DRAM under a linearly varying bitline voltage $V_\mathrm{b}(t)$.}
    \label{fig:linear-response-Vb-dram}
\end{figure}
\begin{figure}
    \centering
    \includegraphics[width=6cm]{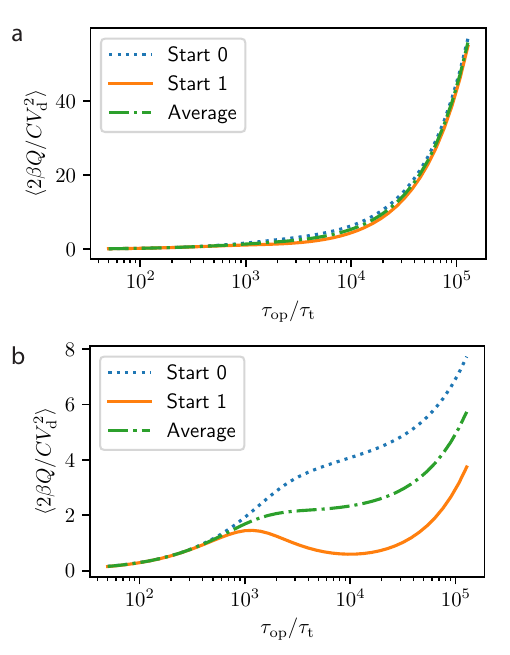}
    \caption{Dissipated heat for SRAM, linearly varying bitline voltage $V_\mathrm{b}(t)$ for driving voltages (a) $V_\mathrm{d}=3$ and (b) $V_\mathrm{d}=6$.}
    \label{fig:linear-response-Vb-sram}
\end{figure}

For the bitline voltage $V_\mathrm{b}$, the friction tensor is uniform and thus the optimal protocol is a linear ramp between the desired parameters. In the long time limit, manipulating the bitline voltage is analogous to manipulating the equilibrium position of a harmonic potential, and this operation is well within the linear response regime. We measured dissipation for a protocol smoothly varying from $V_\mathrm{b}(0^+)=V_\mathrm{d}$ to $V_\mathrm{b}(\tau_\mathrm{op}^-)=0$ in Fig.~\ref{fig:linear-response-Vb-dram}. At short times below the characteristic relaxation time of the system, $\tau_\mathrm{op} < \tau_\mathrm{RC}$, dissipation grows as the capacitor relaxes to equilibrium with the bitline electrode. Starting in the logical 1 state, there is a maximum dissipated heat near the characteristic relaxation time $\tau_\mathrm{op} \approx \tau_\mathrm{RC} \sim 10^3$ before reaching a quasistatic, reversible limit. Our choice of protocol going from $V_\mathrm{b}(0^+)=V_\mathrm{d}$ to $V_\mathrm{b}(\tau_\mathrm{op}^-)=0$ allows the capacitor starting in the logical 1 state to remain in equilibrium with the bitline electrode throughout the protocol, minimizing dissipation in the long-time limit. However, this protocol is necessarily suboptimal starting in the logical 0 state, as the capacitor must first be charged to reach equilibrium with the bitline electrode before relaxing back to the 0 state at long times. This charging cost is reflected in the asymptotic behavior as $Q\rightarrow CV_\mathrm{d}^2/2$. Because we require a single protocol for either starting state, the average behavior reflects both contributions. It is important to note that we instantaneously perturb the bitline voltage from its boundary conditions at $V_\mathrm{b}(0)= V_\mathrm{d}/2$ to $V_\mathrm{b}(0^+)=V_\mathrm{d}$, and the same for $V_\mathrm{b}(\tau_\mathrm{op}^-)=0$ to $V_\mathrm{b}(\tau_\mathrm{op})=V_\mathrm{d}/2$. The original formulation in Ref. \cite{sivak_thermodynamic_2012} does not allow for such a ``bang-bang" protocol, but we may still observe the intervening behavior in the linear response regime.

We have also explored analogous protocols for SRAM, whose behavior changes as a function of driving voltage. For a small driving voltage $V_\mathrm{d}=3$ in Fig.~\ref{fig:linear-response-Vb-sram}(a), housekeeping heat dominates over variations due to the control protocol as thermal noise is close to the scale of driven currents in the system. For a sufficiently high driving voltage $V_\mathrm{d}=6$ in Fig.~\ref{fig:linear-response-Vb-sram}(b), we observe similar behavior as DRAM at intermediate operation times up to $\tau_\mathrm{op} \sim 10^4$, where the control is slow enough that the system can respond to it semi-reversibly. As the driving voltage increases, SRAM reverts to an equilibrium-like behavior when activity is low, as transitions are sufficiently high-energy that they are unlikely. However, housekeeping heat still dominates in the long-time limit. 

\bibliography{references2}

@article{basile_learning_2024,
	title = {Optimal control of static {RAM} erasure: arbitrarily fast operation with finite dissipation},
	volume = {27},
	issn = {1367-2630},
	shorttitle = {Optimal control of static {RAM} erasure},
	doi = {10.1088/1367-2630/ae0ea6},
	number = {10},
	journal = {New Journal of Physics},
	author = {Basile, Tom{\'a}s and Proesmans, Karel},
	month = oct,
	year = {2025},
	pages = {104601},
}

@article{whitelam2024nonequilibrium,
  title={Nonequilibrium formulation of varying-temperature bit erasure},
  author={Whitelam, Stephen},
  journal={Journal of Statistical Mechanics: Theory and Experiment},
  volume={2024},
  number={6},
  pages={063203},
  year={2024},
  publisher={IOP Publishing},
  doi = {10.1088/1742-5468/ad526f}
}

@article{berut_experimental_2012,
  title = {Experimental verification of {Landauer}'s principle linking information and thermodynamics},
  author = {{B{\'e}rut}, Antoine and Arakelyan, Artak and Petrosyan, Artyom and Ciliberto, Sergio and Dillenschneider, Raoul and Lutz, Eric},
  journal = {Nature},
  volume = {483},
  number = {7388},
  pages = {187--189},
  year = {2012},
  publisher = {Nature Publishing Group},
  doi = {10.1038/nature10872}
}

@article{boyd2022shortcuts,
  title = {Shortcuts to thermodynamic computing: The cost of fast and faithful information processing},
  author = {Boyd, Alexander B. and Patra, Ayoti and Jarzynski, Christopher and Crutchfield, James P.},
  journal = {Journal of Statistical Physics},
  volume = {187},
  number = {2},
  pages = {17},
  year = {2022},
  publisher = {Springer},
  doi = {10.1007/s10955-022-02895-9}
}

@article{brillouin_can_1950,
  title = {Can the Rectifier Become a Thermodynamical Demon?},
  author = {Brillouin, L.},
  journal = {Physical Review},
  volume = {78},
  number = {5},
  pages = {627--628},
  year = {1950},
  publisher = {American Physical Society},
  doi = {10.1103/PhysRev.78.627.2}
}

@article{dago_information_2021,
  title = {Information and Thermodynamics: Fast and Precise Approach to {Landauer}'s Bound in an Underdamped Micromechanical Oscillator},
  author = {Dago, Salamb{\^o} and Pereda, Jorge and Barros, Nicolas and Ciliberto, Sergio and Bellon, Ludovic},
  journal = {Physical Review Letters},
  volume = {126},
  number = {17},
  pages = {170601},
  year = {2021},
  publisher = {American Physical Society},
  doi = {10.1103/PhysRevLett.126.170601}
}

@article{engel_optimal_2023,
  title = {Optimal Control of Nonequilibrium Systems through Automatic Differentiation},
  author = {Engel, Megan C. and Smith, Jamie A. and Brenner, Michael P.},
  journal = {Physical Review X},
  volume = {13},
  number = {4},
  pages = {041032},
  year = {2023},
  publisher = {American Physical Society},
  doi = {10.1103/PhysRevX.13.041032}
}

@techreport{energy_ai_2025,
  title = {Energy and {AI}},
  author = {{International Energy Agency}},
  institution = {International Energy Agency},
  year = {2025},
  url = {https://www.iea.org/reports/energy-and-ai}
}

@article{esposito_three_2010,
  title = {Three Detailed Fluctuation Theorems},
  author = {Esposito, Massimiliano and {Van den Broeck}, Christian},
  journal = {Physical Review Letters},
  volume = {104},
  number = {9},
  pages = {090601},
  year = {2010},
  publisher = {American Physical Society},
  doi = {10.1103/PhysRevLett.104.090601}
}

@article{freitas_stochastic_2021,
  title = {Stochastic Thermodynamics of Nonlinear Electronic Circuits: A Realistic Framework for Computing Around {$kT$}},
  author = {Freitas, Nahuel and Delvenne, Jean-Charles and Esposito, Massimiliano},
  journal = {Physical Review X},
  volume = {11},
  number = {3},
  pages = {031064},
  year = {2021},
  publisher = {American Physical Society},
  doi = {10.1103/PhysRevX.11.031064}
}

@article{gao_principles_2021,
  title = {Principles of low dissipation computing from a stochastic circuit model},
  author = {Gao, Chloe Ya and Limmer, David T.},
  journal = {Physical Review Research},
  volume = {3},
  number = {3},
  pages = {033169},
  year = {2021},
  publisher = {American Physical Society},
  doi = {10.1103/PhysRevResearch.3.033169}
}

@article{gillespie_general_1976,
  title = {A general method for numerically simulating the stochastic time evolution of coupled chemical reactions},
  author = {Gillespie, Daniel T},
  journal = {Journal of Computational Physics},
  volume = {22},
  number = {4},
  pages = {403--434},
  year = {1976},
  publisher = {Elsevier},
  doi = {10.1016/0021-9991(76)90041-3}
}

@article{gillespie_stochastic_2007,
  title = {Stochastic Simulation of Chemical Kinetics},
  author = {Gillespie, Daniel T.},
  journal = {Annual Review of Physical Chemistry},
  volume = {58},
  number = {1},
  pages = {35--55},
  year = {2007},
  publisher = {Annual Reviews},
  doi = {10.1146/annurev.physchem.58.032806.104637}
}

@article{giorgini2023thermodynamic,
  title = {Thermodynamic cost of erasing information in finite time},
  author = {Giorgini, Ludovico Theo and Eichhorn, Ralf and Das, M and Moon, W and Wettlaufer, JS},
  journal = {Physical Review Research},
  volume = {5},
  number = {2},
  pages = {023084},
  year = {2023},
  publisher = {American Physical Society},
  doi = {10.1103/PhysRevResearch.5.023084}
}

@article{gu_counting_2020,
  title = {Counting statistics and microreversibility in stochastic models of transistors},
  author = {Gu, Jiayin and Gaspard, Pierre},
  journal = {Journal of Statistical Mechanics: Theory and Experiment},
  volume = {2020},
  number = {10},
  pages = {103206},
  year = {2020},
  publisher = {IOP Publishing},
  doi = {10.1088/1742-5468/abbcd5}
}

@article{helms_stochastic_2025,
  title = {Stochastic thermodynamic bounds on logical circuit operation},
  author = {Helms, Phillip and Chen, Songela W. and Limmer, David T.},
  journal = {Physical Review E},
  volume = {111},
  number = {3},
  pages = {034110},
  year = {2025},
  publisher = {American Physical Society},
  doi = {10.1103/PhysRevE.111.034110}
}

@software{jax2018github,
  title = {{JAX}: composable transformations of {P}ython+{N}um{P}y programs},
  author = {Bradbury, James and Frostig, Roy and Hawkins, Peter and Johnson, Matthew James and Leary, Chris and Maclaurin, Dougal and Necula, George and Paszke, Adam and VanderPlas, Jake and Wanderman-Milne, Skye and Zhang, Qiao},
  version = {0.3.13},
  year = {2018},
  url = {http://github.com/jax-ml/jax}
}

@phdthesis{kidger2021on,
  title = {{O}n {N}eural {D}ifferential {E}quations},
  author = {Kidger, Patrick},
  year = {2021},
  school = {University of Oxford}
}

@misc{kingma_adam_2017,
  title = {Adam: {A} {Method} for {Stochastic} {Optimization}},
  author = {Kingma, Diederik P. and Ba, Jimmy},
  year = {2017},
  eprint = {1412.6980},
  archivePrefix = {arXiv},
  primaryClass = {cs},
  url = {http://arxiv.org/abs/1412.6980},
  doi = {10.48550/arXiv.1412.6980}
}

@article{koomey2010implications,
  title = {Implications of historical trends in the electrical efficiency of computing},
  author = {Koomey, Jonathan and Berard, Stephen and Sanchez, Marla and Wong, Henry},
  journal = {IEEE Annals of the History of Computing},
  volume = {33},
  number = {3},
  pages = {46--54},
  year = {2010},
  publisher = {IEEE},
  doi = {10.1109/MAHC.2010.28}
}

@article{landauer_irreversibility_1961,
  title = {Irreversibility and {Heat} {Generation} in the {Computing} {Process}},
  author = {Landauer, Rolf},
  journal = {IBM Journal of Research and Development},
  volume = {5},
  number = {3},
  pages = {183--191},
  year = {1961},
  publisher = {IBM},
  doi = {10.1147/rd.53.0183}
}

@book{limmer2024statistical,
  title = {Statistical Mechanics and Stochastic Thermodynamics},
  author = {Limmer, David T},
  year = {2024},
  publisher = {Oxford University Press}
}

@article{PhysRevE.102.032105,
  title = {Optimal finite-time bit erasure under full control},
  author = {Proesmans, Karel and Ehrich, Jannik and Bechhoefer, John},
  journal = {Physical Review E},
  volume = {102},
  issue = {3},
  pages = {032105},
  numpages = {12},
  year = {2020},
  publisher = {American Physical Society},
  doi = {10.1103/PhysRevE.102.032105}
}

@article{proesmans_finite-time_2020,
  title = {Finite-{Time} {Landauer} {Principle}},
  author = {Proesmans, Karel and Ehrich, Jannik and Bechhoefer, John},
  journal = {Physical Review Letters},
  volume = {125},
  number = {10},
  pages = {100602},
  year = {2020},
  publisher = {American Physical Society},
  doi = {10.1103/PhysRevLett.125.100602}
}

@article{sagawa_minimal_2009,
  title = {Minimal {Energy} {Cost} for {Thermodynamic} {Information} {Processing}: {Measurement} and {Information} {Erasure}},
  author = {Sagawa, Takahiro and Ueda, Masahito},
  journal = {Physical Review Letters},
  volume = {102},
  number = {25},
  pages = {250602},
  year = {2009},
  publisher = {American Physical Society},
  doi = {10.1103/PhysRevLett.102.250602}
}

@article{sarpeshkar2002white,
  title = {White noise in {MOS} transistors and resistors},
  author = {Sarpeshkar, Rahul and Delbruck, Tobias and Mead, Carver A},
  journal = {IEEE Circuits and Devices Magazine},
  volume = {9},
  number = {6},
  pages = {23--29},
  year = {2002},
  publisher = {IEEE},
  doi = {10.1109/MCD.2002.1035347}
}

@article{seifert_stochastic_2012,
  title = {Stochastic thermodynamics, fluctuation theorems and molecular machines},
  author = {Seifert, Udo},
  journal = {Reports on Progress in Physics},
  volume = {75},
  number = {12},
  pages = {126001},
  year = {2012},
  publisher = {IOP Publishing},
  doi = {10.1088/0034-4885/75/12/126001}
}

@article{shimizu_thermodynamic_2025,
	title = {Thermodynamic {Constraints} in {Dynamic} {Random}-{Access} {Memory} {Cells}: {Experimental} {Verification} of {Energy} {Efficiency} {Limits} in {Information} {Erasure}},
	volume = {136},
	issn = {0031-9007, 1079-7114},
	shorttitle = {Thermodynamic {Constraints} in {Dynamic} {Random}-{Access} {Memory} {Cells}},
	doi = {10.1103/1sgm-dhys},
	number = {11},
	journal = {Physical Review Letters},
	author = {Shimizu, Takase and Chida, Kensaku and Yamahata, Gento and Nishiguchi, Katsuhiko},
	month = mar,
	year = {2026},
	pages = {117103},
}

@article{sivak_thermodynamic_2012,
  title = {Thermodynamic {Metrics} and {Optimal} {Paths}},
  author = {Sivak, David A. and Crooks, Gavin E.},
  journal = {Physical Review Letters},
  volume = {108},
  number = {19},
  pages = {190602},
  year = {2012},
  publisher = {American Physical Society},
  doi = {10.1103/PhysRevLett.108.190602}
}

@techreport{SPICE,
  title = {{SPICE} ({S}imulation {P}rogram with {I}ntegrated {C}ircuit {E}mphasis)},
  author = {Nagel, Laurence W. and Pederson, D.O.},
  institution = {University of California, Berkeley},
  number = {UCB/ERL M382},
  year = {1973},
  url = {http://www2.eecs.berkeley.edu/Pubs/TechRpts/1973/22871.html}
}

@book{taur_ning,
  title = {Fundamentals of Modern VLSI Devices},
  author = {Taur, Yuan and Ning, Tak H.},
  year = {1998},
  publisher = {Cambridge University Press}
}

@inproceedings{van_brandt_non-landauer_2023,
  title = {The non-{Landauer} {Bound} for the {Dissipation} of {Bit} {Writing} {Operation}},
  author = {{Van Brandt}, L{\'e}opold and Delvenne, Jean-Charles},
  booktitle = {2023 {IEEE} 23rd {International} {Conference} on {Nanotechnology} ({NANO})},
  pages = {726--731},
  year = {2023},
  publisher = {IEEE},
  address = {Jeju City, Korea, Republic of},
  doi = {10.1109/NANO58406.2023.10231222}
}

@article{van_brandt_noisedissipation_2023,
  title = {Noise--dissipation relation for nonlinear electronic circuits},
  author = {{Van Brandt}, L{\'e}opold and Delvenne, Jean-Charles},
  journal = {Applied Physics Letters},
  volume = {122},
  number = {26},
  pages = {263507},
  year = {2023},
  publisher = {AIP Publishing},
  doi = {10.1063/5.0152883}
}

@article{wimsatt_refining_2021,
  title = {Refining {Landauer}'s {Stack}: {Balancing} {Error} and {Dissipation} {When} {Erasing} {Information}},
  author = {Wimsatt, Gregory W. and Boyd, Alexander B. and Riechers, Paul M. and Crutchfield, James P.},
  journal = {Journal of Statistical Physics},
  volume = {183},
  number = {1},
  pages = {16},
  year = {2021},
  publisher = {Springer},
  doi = {10.1007/s10955-021-02733-1}
}

@article{zulkowski_optimal_2014,
  title = {Optimal finite-time erasure of a classical bit},
  author = {Zulkowski, Patrick R. and DeWeese, Michael R.},
  journal = {Physical Review E},
  volume = {89},
  number = {5},
  pages = {052140},
  year = {2014},
  publisher = {American Physical Society},
  doi = {10.1103/PhysRevE.89.052140}
}

@software{code,
  author={Chen, Songela W. and Limmer, David T.},
  title={Code and Data for "Optimal control of bit erasure in stochastic random access memory"},
  year=2026,
  publisher={Zenodo},
  doi={10.5281/zenodo.19582802},
}

@article{ciampini_erasure_2025,
	title = {Erasure of a nonequilibrium memory beyond {Landauer}'s bound using levitated optomechanics with spatio-temporal optical control},
	volume = {7},
	issn = {2643-1564},
	doi = {10.1103/r81x-zblx},
	number = {4},
	journal = {Physical Review Research},
	author = {Ciampini, Mario A. and Wenzl, Tobias and Konopik, Michael and Thalhammer-Thurner, Gregor and Aspelmeyer, Markus and Lutz, Eric and Kiesel, Nikolai},
	month = dec,
	year = {2025},
	pages = {043321},
}

@article{dago_reliability_2024,
	title = {Reliability and {Operation} {Cost} of {Underdamped} {Memories} during {Cyclic} {Erasures}},
	volume = {3},
	copyright = {© 2023 The Authors. Advanced Physics Research published by Wiley-VCH GmbH},
	issn = {2751-1200},
	doi = {10.1002/apxr.202300074},
	number = {2},
	journal = {Advanced Physics Research},
	author = {Dago, Salambô and Ciliberto, Sergio and Bellon, Ludovic},
	year = {2024},
	pages = {2300074},
}

@article{freitas_reliability_2022,
	title = {Reliability and entropy production in nonequilibrium electronic memories},
	volume = {105},
	url = {https://doi.org/10.1103/PhysRevE.105.034107},
	number = {3},
	journal = {Physical Review E},
	author = {Freitas, Nahuel and Proesmans, Karel and Esposito, Massimiliano},
	month = mar,
	year = {2022},
	pages = {034107},
}

@misc{murphy_dissipation-reliability_2026,
	title = {Dissipation-{Reliability} {Tradeoff} for {Stochastic} {CMOS} {Bits} in {Series}},
	doi = {10.48550/arXiv.2603.04658},
	urldate = {2026-03-15},
	publisher = {arXiv},
	author = {Murphy, Cathryn and Nicholson, Schuyler and Freitas, Nahuel and Penocchio, Emanuele and Gingrich, Todd},
	month = mar,
	year = {2026},
	note = {arXiv:2603.04658 [cond-mat]},
}

\end{document}